\documentclass[preprint]{emulateapj}

\shortauthors{Koyamatsu et al.}

\usepackage{aas_macros}
\usepackage{amsmath}
\usepackage[breaklinks=true,colorlinks=true,linkcolor=blue,citecolor=blue,urlcolor=red,
setpagesize=false]{hyperref}

\begin{document}

\title{The Disappearing Envelope around the Transitional Class~I Object L43}


\author{Shin Koyamatsu\altaffilmark{1,2,3},
Shigehisa Takakuwa\altaffilmark{4},
Masahiko Hayashi\altaffilmark{5},
Satoshi Mayama\altaffilmark{6},
Nagayoshi Ohashi\altaffilmark{2}}

\altaffiltext{1}{Department of Astronomy, 
Graduate School of Science, The University of Tokyo, 
Hongo 7-3-1, Bunkyo-ku, Tokyo 113-0033, Japan}
\altaffiltext{2}{Subaru Telescope,
National Astronomical Observatory of Japan,
650 North A'ohoku Place, Hilo, HI 96720, USA}
\altaffiltext{3}{e-mail: shin.koyamatsu@nao.ac.jp}
\altaffiltext{4}{Academia Sinica Institute of Astronomy and Astrophysics,
P.O. Box 23-141, Taipei 10617, Taiwan}
\altaffiltext{5}{National Astronomical Observatory of Japan,
Osawa 2-21-1, Mitaka-shi, Tokyo 181-8588, Japan}
\altaffiltext{6}{Graduate University for Advanced Studies,
Shonan International Village,
Hayama-cho, Miura-gun, Kanagawa 240-0193, Japan}

\begin{abstract}
 We present SMA interferometric observations of the $^{12}$CO ($J=2\text{--}1$), $^{13}$CO ($J=2\text{--}1$), and C$^{18}$O ($J=2\text{--}1$) lines and 225 GHz continuum emission and SMT single-dish observations of C$^{18}$O ($J=2\text{--}1$) toward L43, a protostellar object in transition from Class I to II. The 225 GHz continuum emission shows a weak ($\sim 23.6\,\mathrm{mJy}$), compact ($< 1000\,\mathrm{AU}$) component associated with the central protostar. Our simulated observations show that it can be explained by dust thermal emission arising from an envelope which has a hole or a constant intensity region within a few hundred AU of the protostar. This suggests the disappearance or a lower concentration distribution of the envelope on a small scale. The $^{12}$CO and $^{13}$CO emission exhibit molecular outflows to the south and north. The C$^{18}$O emission shows two molecular blobs, which correspond to the reflection nebulosity seen in near-infrared images, while there is no C$^{18}$O emission associated with the protostar. The near-infrared features are likely due to the scattering at the positions of the blobs. The visible scattering features should result from the optical thinness of the envelope material, which is consistent with the less-concentrated distribution in the continuum emission. From single-dish observations, we found that the mass of the envelope ($\sim 1.5\,\mathrm{M_{\odot}}$) $+$ protostar ($\sim 0.5\,\mathrm{M_{\odot}}$) is comparable with the virial mass of $M_\mathrm{vir}=1.0\,\mathrm{M_{\odot}}$ within $40\arcsec$. This suggests that the envelope is likely gravitationally bound. We suggest that the protostellar envelope of L43 has been disappearing by consumption through accretion, at least in the close vicinity of the protostar.
\end{abstract}

 \section{Introduction}
 Observational and theoretical studies in the past decades have revealed that low-mass protostars (Class 0 and Class I sources) evolve into T~Tauri stars (Class II sources) surrounded by protoplanetary disks. A low-mass accreting protostar is associated with an envelope or a dense molecular cloud core, with a mass typically a few times the final stellar mass and a size of $1,000\text{--}10,000\,\mathrm{AU}$, while an optically visible T~Tauri star is accompanied by a protoplanetary disk with a typical mass of $0.1\text{--}1\%$ of the star and a size of $100\text{--}1,000\,\mathrm{AU}$ \citep[e.g.][]{andre00,myers00,andrews05}. Although a protostellar envelope is thought to disappear in part by accretion onto the central star-disk system, and in part by being blown away by the associated outflow in the course of evolution from a protostar to a T~Tauri star \citep{nakano95}, details of the disappearing processes are still poorly understood. Observations of a transitional object from an embedded protostar to a T~Tauri star is crucial to understand the mechanism for disappearing protostellar envelope.

 There are several low-mass young stars classified as T~Tauri stars but associated with relatively large amounts of material in their envelopes. These transitional objects are characterized by a flat spectral energy distribution (SED) in infrared wavelengths \citep[$2.2\text{--}10\,\mathrm{\mu m}$;][]{greene94}, although they are usually classified either as Class~I or Class~II according to their near to mid-infrared SEDs. HL~Tau is an archetypal flat-spectrum object with a flattened infalling envelope \citep{hayashi93}, while DG~Tau, another representative flat-spectrum source, shows an expanding envelope \citep{kitamura96}. Each envelope has a similar mass of $0.03\,\mathrm{M_{\odot}}$, slightly higher than the typical mass of many protoplanetary disks. Another prototypical flat-spectrum object, T~Tau, still possesses $0.31\text{--}1.3\,\mathrm{M_{\odot}}$ of surrounding gas \citep{momose96}. The structure and kinematics around the transitional objects appear to be complicated, and require further studies.

 L43, also called RNO~91, is one of several such transitional objects located in the Ophiuchus molecular cloud complex at a distance of $\sim 125\,\mathrm{pc}$ \citep{de-geus90}. The source has a bolometric luminosity of $4.3\,\mathrm{L_{\odot}}$ \citep{terebey93} and a central stellar mass of $0.5\,\mathrm{M_{\odot}}$ \citep{levreault88}. L43 used to be classified as a Class II object, and its bolometric temperature was estimated to be $715\,\mathrm{K}$ \citep{andre94, chen95}. Recent data, however, show that L43 is now classified as a Class I object with a lower bolometric temperature of $337.6\,\mathrm{K}$, which is still higher than that of typical Class I objects. Single-dish observations of L43 in the $850\,\mathrm{\mu m}$ continuum emissions showed the presence of an extended ($\sim 7,000\,\mathrm{AU}$) dusty envelope \citep{shirley00}, and interferometric mosaicking observations of L43 unveiled the conspicuous $U$-shaped, large-scale ($> 50,000\,\mathrm{AU}$) blueshifted molecular outflow toward the south-east from the driving source with a position angle of $155\degr$ \citep{lee02,lee05}. Single-field interferometric observations of the close vicinity ($< 10,000\,\mathrm{AU}$) of L43 also showed the outflow-like structure but with the orientation toward the south rather than south-east\citep{arce06}. Near-infrared images of L43 taken with the Subaru Telescope exhibit reflection nebulosity and extinction maps which might be attributed to the disk-like structure around the central star \citep{mayama07}. Optical polarization observations also suggest the presence of the circumstellar disk around L43 \citep{heyer90,scarrott93}. These results show that L43 is one of the ideal objects in which to study the transition from the protostellar stage to the T~Tauri stage. Table~\ref{tab:l43} summarizes the properties of L43.
\begin{deluxetable}{lcccc}
\tabletypesize{\scriptsize}
\tablecaption{
Properties of L43
\label{tab:l43}
}
\tablewidth{0pt}
\tablehead{
\colhead{} &
\colhead{Distance\tablenotemark{a}} &
\colhead{$L_\mathrm{bol}$\tablenotemark{b}} &
\colhead{$T_\mathrm{bol}$\tablenotemark{c}} &
\colhead{$M_\mathrm{star}$\tablenotemark{d}} \\
\colhead{Name} &
\colhead{(pc)} &
\colhead{(L$_{\odot}$)} &
\colhead{(K)} &
\colhead{(M$_{\odot}$)}
}
\startdata
L43 (RNO~91) & 125 & $2.5\pm 0.1$ & $337.6\pm 18.9$ & 0.5
\enddata
\tablenotetext{a}{\citet{de-geus90}.}
\tablenotetext{b}{Bolometric luminosity \citep{chen09}.}
\tablenotetext{c}{Bolometric temperature \citep{chen09}.}
\tablenotetext{d}{Central stellar mass \citep{levreault88}.}
\end{deluxetable}

In this paper, we present results of the SubMillimeter Array (SMA)\footnote{The Submillimeter Array (SMA) is a joint project between the Smithsonian Astrophysical Observatory and the Academia Sinica Institute of Astronomy and Astrophysics and is funded by the Smithsonian Institution and the Academia Sinica.} observations of L43 in $^{12}$CO ($J=2\text{--}1$), $^{13}$CO ($J=2\text{--}1$), and C$^{18}$O ($J=2\text{--}1$) lines and 225~GHz continuum emission, together with the 10~m Submillimeter Telescope (SMT) observations in C$^{18}$O ($J=2\text{--}1$) line. Our analyses of the data have shown details of the molecular outflow, the dispersion of the protostellar envelope, and the circumstellar disk. We describe below the SMA and SMT observations ($\S$~2) and continuum and molecular-line results ($\S$~3). In the last section ($\S$~4), we will discuss the dust opacity index, the origin of the dust emission, and the physical condition of the protostellar envelope in L43. 

 \section{Observations}\label{sec:obs}
 L43 was observed with the compact configuration of the SMA on 2005 June 28 in $^{12}$CO ($J=2\text{--}1$; 230.538000~GHz), $^{13}$CO ($J=2\text{--}1$; 220.398684~GHz), and C$^{18}$O ($J=2\text{--}1$; 219.560358~GHz) lines, and in 1.3~mm continuum emission simultaneously. Details of the SMA are described by \citet{ho04}. For the molecular-line observations, 128 spectral channels were assigned to spectral windows of the SMA correlator (``chunks'') with a total bandwidth of 82\ MHz, providing a frequency resolution of 812~kHz, which corresponds to a velocity resolution of $\sim 1.06\,\mathrm{km\,s^{-1}}$ at the C$^{18}$O ($J=2\text{--}1$) frequency. In each sideband there was a total of 24 chunks, and all the chunks at both sidebands (except for those assigned to the $^{12}$CO, $^{13}$CO and C$^{18}$O lines) were combined to make a single continuum channel. The effective continuum bandwidth is $\sim 3.8\,\mathrm{GHz}$. In the following, we refer to the continuum emission taken with the present SMA observations as ``225~GHz continuum emission'', since the LO frequency was $\sim 225.10\,\mathrm{GHz}$. The minimum projected baseline length was $\sim 7.3\,\mathrm{k\lambda}$, and in the case of the Gaussian emission distribution with an FWHM size of $\sim 23\arcsec$ ($\sim 2,900\,\mathrm{AU}$) the peak flux that can be recovered $\sim 10\%$ of the original peak flux \citep{wilner94}. The estimated uncertainty in the absolute flux calibration was $\sim 30\%$. The raw visibility data were calibrated and flagged with MIR, which is an IDL-based data reduction package \citep{scoville93}. After the calibrations, we constructed $^{12}$CO, $^{13}$CO, C$^{18}$O, and 225~GHz continuum maps with the natural weighting using the MIRIAD software package \citep{sault95}. We did not apply primary beam correction, since we focused mainly on compact structures in the vicinity of the central star of L43 in the SMA images.

We performed single-dish mapping observations in the C$^{18}$O ($J=2\text{--}1$) line toward L43 with the Submillimeter Telescope (SMT) on 2012 February 3--5. A total of 249 positions centered on $\alpha(\mathrm{J2000})=16^\mathrm{h} 34^\mathrm{m} 29^\mathrm{s}.3$, $\delta(\mathrm{J2000})=-15\degr 47\arcmin 01\farcs 5$ were mapped with a grid spacing of $20\arcsec$. The mapped areas are $6.3\arcmin\times 3.0\arcmin$ covering the area south of the central star and $4.3\arcmin\times 2.0\arcmin$ covering the area north, as indicated in Figure~\ref{fig:c18osmt}. The integration time for each pointing (on and off) was $30\,\mathrm{s}$, and each position in the map was pointed to at least twice. The central position was adopted to check the relative flux calibration, and flux uncertainty was estimated to be $\sim 10\text{--}20\%$. The telescope pointing was checked at the beginning of the four-hour observations by observing planets. The antenna temperature $T_\mathrm{a}^{*}$ was converted to the brightness temperature $T_\mathrm{mb}$ by $T_\mathrm{mb}=T_\mathrm{a}^{*}/\eta_\mathrm{mb}$, where $\eta_\mathrm{mb}=0.74$ is the main beam efficiency of the SMT. The details of observations are summarized in Tables~\ref{tab:smaobs}--\ref{tab:smtobs}.
\begin{deluxetable}{lc}
\tabletypesize{\scriptsize}
\tablecaption{
Characteristics of the SMA line and continuum maps
\label{tab:smaobs}
}
\tablewidth{0pt}
\tablehead{
\colhead{Parameter} & \colhead{Value}
}
\startdata
Date & 2005 June 28 \\
Right Ascension  (J2000.0) & $16^\mathrm{h} 34^\mathrm{m} 29^\mathrm{s}.33$ \\
Declination  (J2000.0) & $-15\degr 47\arcmin 01\farcs 5$ \\
Number of Antennas & 7 \\
Primary Beam HPBW & $\sim 56\arcsec$ \\
Baseline Coverage & $7.3\text{--}160\,\mathrm{k\lambda}$ \\
Frequency Resolution & $812\,\mathrm{kHz}$ ($\sim 1.06\,\mathrm{km\,s^{-1}}$) \\
Bandwidth & $3.8\,\mathrm{GHz}$ \\
Flux Calibrator & Callisto \\
Gain Calibrators & 1743-038, 1733-130 \\
Flux (1743-038) & $1.87\,\mathrm{Jy}$ (Upper)\tablenotemark{a}, $1.93\,\mathrm{Jy}$ (Lower)\tablenotemark{a} \\
Flux (1733-130) & $1.14\,\mathrm{Jy}$ (Upper)\tablenotemark{a}, $1.19\,\mathrm{Jy}$ (Lower)\tablenotemark{a} \\
Passband Calibrator & 3c454.3 \\
System Temperature (DSB) & $\sim 300\text{--}500\,\mathrm{K}$
\enddata
\tablenotetext{a}{``Upper'' and ``lower'' denote the flux densities of the calibrator at the upper and lower sidebands, respectively.}
\end{deluxetable}
\begin{deluxetable*}{llccc}
\tabletypesize{\scriptsize}
\tablecaption{
Characteristics of the line and continuum maps
\label{tb:map}
}
\tablewidth{0pt}
\tablehead{
\colhead{} &
\colhead{Frequency} &
\colhead{Synthesized Beam HPBW} &
\colhead{rms Noise Level} &
\colhead{Conversion Factor} \\
\colhead{Line/Continuum} &
\colhead{(GHz)} &
\colhead{(arcsec)} &
\colhead{($\mathrm{Jy\,beam^{-1}}$)} &
\colhead{($\mathrm{K / (Jy\,beam^{-1})}$)}
}
\startdata
$^{12}$CO ($J=2\text{--}1$) & 230.538000 & $4\farcs 4\times 3\farcs 3$
($\text{P.A.}=19\degr$) & 0.169 & 4.89 \\
$^{13}$CO ($J=2\text{--}1$) & 220.398684 & $4\farcs 3\times 3\farcs 4$
($\text{P.A.}=16\degr$) & 0.140 & 5.59 \\
C$^{18}$O ($J=2\text{--}1$) & 219.560358 & $4\farcs 4\times 3\farcs 5$
($\text{P.A.}=15\degr$) & 0.139 & 5.33 \\
Continuum & 225.10 & $4\farcs 4\times 3\farcs 4$
($\text{P.A.}=17\degr$) & $2.38\times 10^{-3}$ & 5.20
\enddata
\end{deluxetable*}
\begin{deluxetable}{lc}
\tabletypesize{\scriptsize}
\tablecaption{
Summary of the SMT observations
\label{tab:smtobs}
}
\tablewidth{0pt}
\tablehead{
\colhead{Parameter} & \colhead{Value}
}
\startdata
Date & 2012 February 3--5 \\
Right Ascension (J2000.0) & $16^\mathrm{h} 34^\mathrm{m} 29^\mathrm{s}.3$ \\
Declination (J2000.0) & $-15\degr 47\arcmin 01\farcs 5$ \\
Beam Size & $\sim 34\farcs 3$ \\
Velocity Resolution & $0.34\,\mathrm{km\,s^{-1}}$ \\
Main Beam Efficiency & $74\%$ \\
Noise Level & $< 100\,\mathrm{mK}$ \\
System Temperature & $\sim 170\text{--}250\,\mathrm{K}$
\enddata
\end{deluxetable}

\section{Results}\label{sec:result}
\subsection{225~GHz Continuum Emission}\label{sec:cont}
\begin{figure}
 \centering
 \includegraphics[bb=0 0 644 500, width=\hsize]{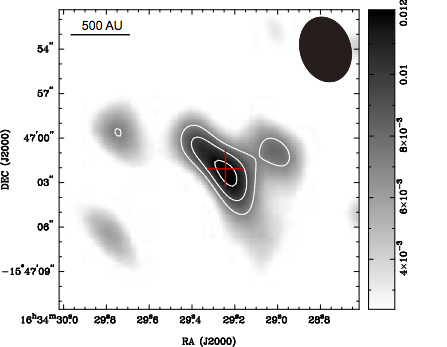}
 \caption{
 225\,GHz continuum image of L43 observed with the SMA. The contour levels are from $3\sigma$ in steps of $1\sigma$ ($1\sigma =2.38\,\mathrm{mJy\,beam^{-1}}$). The highest contour level is $5\sigma$. The cross indicates the peak position of a Gaussian fitting to the continuum emission. The filled elipse at the top right corner shows the synthesized beam ($4\farcs4\times 3\farcs4$; $\text{P.A .}=17\degr$). 
 \label{fig:cnt}
 }
\end{figure}
Figure~\ref{fig:cnt} shows the 225\ GHz continuum map toward L43 observed with the SMA. The continuum emission is clearly detected at above $5\sigma$ level toward the central stellar position. In addition to the main peak located at the central stellar position, there is a secondary peak 3" to the northwest. While the continuum distribution possibly suggests the presence of this and multiple other components in the area, we cannot confirm them due to the low signal-to-noise ratio and will not discuss them further here. The peak and total flux densities of the primary component are $12.1\pm 2.4\,\mathrm{mJy\,beam^{-1}}$ and $23.6\pm 2.4\,\mathrm{mJy}$, respectively. Although the primary continuum component shows apparent elongation from northeast to southwest, a 2-dimensional Gaussian fitting to the image does not provide a beam-deconvolved size and a position angle, presumably due to the low signal-to-noise ratio.

\begin{figure}
 \centering
 \includegraphics[bb=0 0 576 432, width=\hsize]{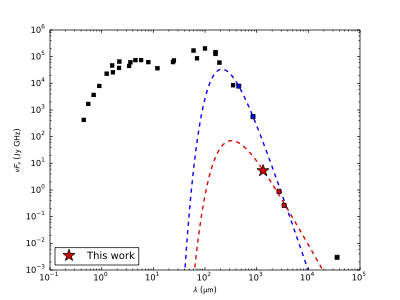}
 \caption{
 The spectral energy distribution of L43. Measurements are primarily taken from the compilation of \citet{chen09}. In addition to this, optical measurements are taken from \citet{myers87}, infrared from \citet{shirley00}, millimeter from \citet{arce06}, and VLA measurement from \citet{anglada02}. Our measurement is presented as the star. The blue and red symbols indicate the measurements at submillimeter and millimeter wavelengths, respectively. The blue and red dashed lines represent dust thermal emission for best fit $\beta$ with $T_\mathrm{d}=10\,\mathrm{K}$ at submillimeter and millimeter wavelengths, respectively. 
 \label{fig:sed}
 }
\end{figure}
 To illustrate the nature of the dust continuum emission, the spectral energy distribution (SED) toward L43 is presented in Figure~\ref{fig:sed}. The SED shows a fairly flat or slightly rising spectrum from near- to far-infrared, with less excess at far-infrared as compared to that of a typical Class I object. This is consistent with the interpretation that L43 is at a transitional phase from a protostar to a T~Tauri star. The three data points at millimeter wavelengths (marked in red in Figure~\ref{fig:sed}) were taken from interferometric observations (including the present work)with beam sizes of $4\text{--}13\arcsec$, while data points at far-infrared were taken from single-aperture observations with $40\text{--}70\arcsec$ beam sizes \citep{chen09}. It is clear that values at millimeter wavelengths are lower than those expected from extrapolation of the measurements at far-infrared wavelengths. This is probably because of artifacts caused by the interferometric observations resolving out extended continuum emission. In other words, the millimeter interferometric observations selectively detected a central part of the circumstellar structure rather than the extended dust emission. We discuss the property and origin of dust continuum emission at millimeter wavelengths in \S~\ref{sec:beta} and \S~\ref{sec:dust}.

\subsection{CO Line Emission}
 Figures~\ref{fig:12coch}--\ref{fig:C18Oblob1} show velocity channel maps and integrated intensity maps of $^{12}$CO ($J=2\text{--}1$), $^{13}$CO ($J=2\text{--}1$), and C$^{18}$O ($J=2\text{--}1$) lines. We adopt the systemic velocity of the protostar L43 as $V_{\rm LSR}=0.52\pm 0.01\,\mathrm{km\,s^{-1}}$, which was derived from the Gaussian fitting to the single-dish DCO$^{+}$ ($J=3\text{--}2$) and C$^{18}$O ($J=2\text{--}1$) spectra toward L43 \citep{chen09}.

\subsubsection{$^{12}$CO ($J=2\text{--}1$) Emission}\label{sec:12co}
\begin{figure*}
 \centering
 \includegraphics[bb=0 0 569 549, width=0.8\hsize]{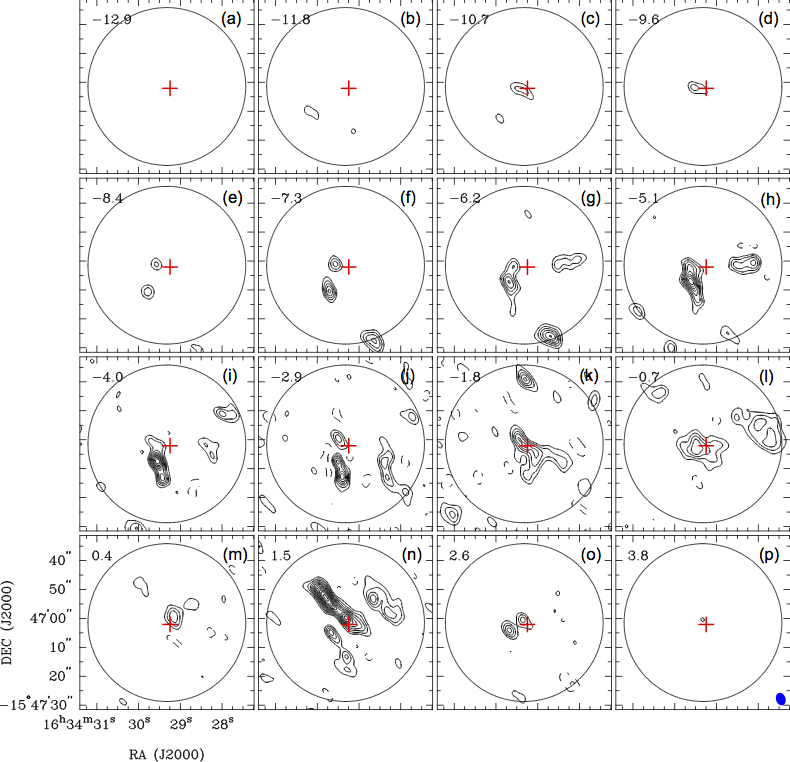}
 \caption{
   Velocity channel maps of the $^{12}$CO ($J=2\text{--}1$) line emission in L43 observed with the SMA.
   The contour levels are from $5\sigma$ in steps of $3\sigma$ ($1\sigma =153\,\mathrm{mJy\,beam^{-1}}$).
   The central velocity of each channel is shown at the top left in units of $\mathrm{km\,s^{-1}}$.
   The crosses indicate the peak position of the 225~GHz continuum emission in Figure~\ref{fig:cnt},
   and the filled ellipse at the bottom right corner in panel $p$ shows the synthesized beam ($4\farcs 4\times 3\farcs 3$; $\mathrm{P.A.}=19\degr$).
   The open circles show the field of view at the frequency of $^{12}$CO ($\sim 54\farcs5$).
   \label{fig:12coch}
 }
\end{figure*}

 The $^{12}$CO ($J=2\text{--}1$) emission was detected at above $5\sigma$ in the velocity range from $V_\mathrm{LSR}= -12.4$ to $4.4\,\mathrm{km\,s^{-1}}$ as shown in Figure~\ref{fig:12coch}. The detected velocity range coincides well with that of the same line obtained with the CSO at an angular resolution almost 10 times as large \citep{chen09}. The single-dish $^{12}$CO ($J=2\text{--}1$) line profile shows a blue-red asymmetry with stronger redshifted emission, which is also evident in the channel maps. The blueshifted CO components (panels $b\text{--}l$) are located at $\sim 3\arcsec$ east, $\sim 6\arcsec$ southeast and $\sim 15\arcsec$ west of the central protostellar position, marked by a cross in each panel. Although we detected possible emission at the edge of the primary beam, it is difficult to interpret because of the complicated structure of this object as shown in the large-scale outflow. On the other hand, the redshifted emission (panels $n$--$p$), predominantly detected at $1.5\,\mathrm{km\,s^{-1}}$, shows a strong feature elongated out from the protostellar position to the northeast. In addition, redshifted emission is also detected at southeast and northwest. At the systemic velocity channel (panel $m$), emission is detected in the vicinity of central protostar, which is weaker than neighboring channels. Although the single-dish observations show a self-absorption feature in $^{12}$CO at the systemic velocity \citep{chen09}, our SMA observations detected even weaker emission than the single-dish observations at the systemic velocity. This is because the $^{12}$CO emission is partly resolved out due to the interferometric observations at the systemic velocity where extended emission is more dominant.

\begin{figure*}
 \centering
 \includegraphics[bb=0 0 1494 567, width=0.8\hsize]{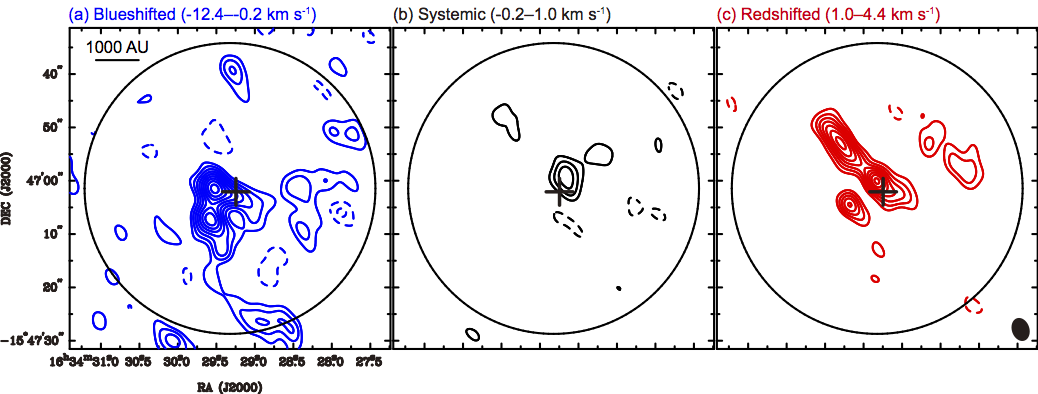}
 \caption{
 Integrated intensity maps of the $^{12}$CO ($J=2\text{--}1$) line emission in L43 observed with the SMA. Each panel shows the integrated intensity map over (a) blueshifted (panels $b\text{--}l$ in Figure~\ref{fig:12coch}), (b) systemic (panel $m$), and (c) redshifted (panels $n\text{--}p$) channels, respectively. The contour levels are from $5\sigma$ in steps of $3\sigma$, where the $1\sigma$ noise levels are $0.563\,\mathrm{Jy\,beam^{-1}\,km\,s^{-1}}$, $0.170\,\mathrm{Jy\,beam^{-1}\,km\,s^{-1}}$, and $0.294\,\mathrm{Jy\,beam^{-1}\,km\,s^{-1}}$, respectively. The crosses indicate the peak position of the 225~GHz continuum emission in Figure~\ref{fig:cnt}, and the filled ellipse at the bottom right corner in panel (c) shows the synthesized beam ($4\farcs 4\times 3\farcs 3$; $\mathrm{P.A.}=19\degr$). The open circles show the field of view at the frequency of $^{12}$CO ($\sim 54\farcs 5$).
 \label{fig:12comom0}
 }
\end{figure*}

 Figure~\ref{fig:12comom0} shows intensity maps of the $^{12}$CO emission integrated over three different velocity ranges; blueshifted ($V_\mathrm{LSR}=-12.4\text{--}-0.2\,\mathrm{km\,s^{-1}}$), systemic ($V_\mathrm{LSR}=-0.2\text{--}1.0\,\mathrm{km\,s^{-1}}$), and redshifted ($V_\mathrm{LSR}=1.0\text{--}4.4\,\mathrm{km\,s^{-1}}$) velocities. At the blueshifted velocity, there appear four main emission components. The closest $^{12}$CO component to the central protostar protrudes eastward, and the second component is located south of the first one. The third and fourth components are located west of the first and second components. Although there are other detections at the edge of SMA primary beam ($22\arcsec$ north, $24\arcsec$ northwest, and $28\arcsec$ south), the blueshifted emission tends to be south from the protostellar position. In the redshifted velocity, the $^{12}$CO emission primarily shows an elongated feature northeast of the protostar. In addition, two weaker features are seen at $\sim 7\arcsec$ southeast and $\sim 15\arcsec$ northwest of the protostar. The redshifted emission appears to be located on the opposite side from the blueshifted one. The $^{12}$CO emission around the systemic velocity appears to be associated with the central protostar. This emission, however, is largely resolved out as mentioned above, and we will not discuss it in this paper. The indication that blueshifted and redshifted components are located at the south and north, respectively, is roughly consistent with the previous studies of the outflow associated with L43 \citep{lee02,lee05,arce06}. We thus consider that the $^{12}$CO ($J=2\text{--}1$) emission selectively traces the molecular outflow driven by L43 in the same manner as typical protostars do.

\subsubsection{$^{13}$CO ($J=2\text{--}1$) Emission}\label{sec:13co}
\begin{figure*}
 \centering
 \includegraphics[bb=0 0 569 549, width=0.8\hsize]{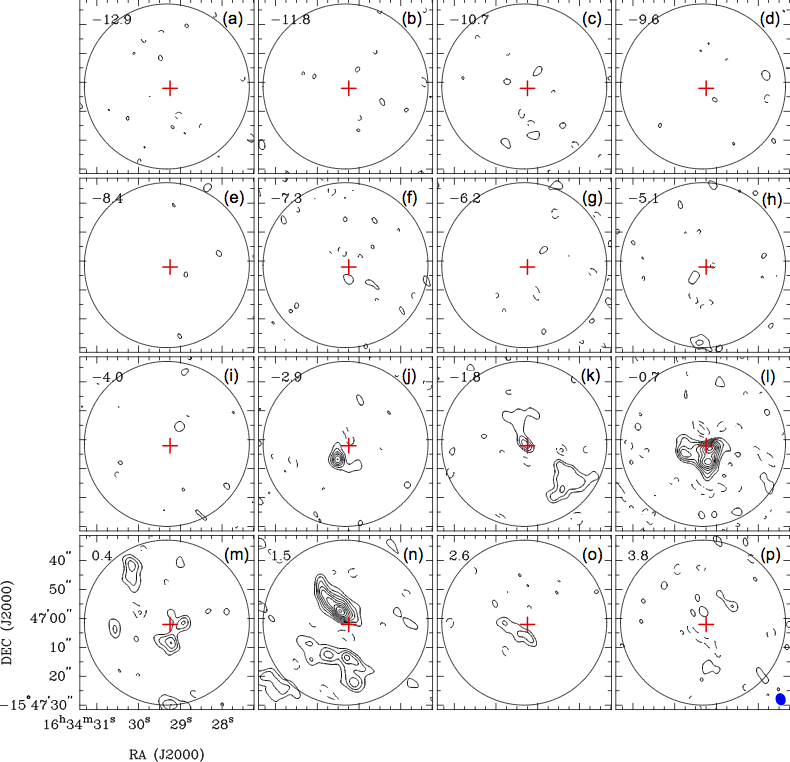}
 \caption{
   Velocity channel maps of the $^{13}$CO ($J=2\text{--}1$) line emission in L43 observed with the SMA.
   The contour levels are from $3\sigma$ in steps of $2\sigma$ ($1\sigma =151\,\mathrm{mJy\,beam^{-1}}$).
   The central velocity of each channel is shown at the top left in units of $\mathrm{km\,s^{-1}}$.
   The crosses indicate the peak position of the 225~GHz continuum emission in Figure~\ref{fig:cnt},
   and the filled ellipse at the bottom right corner in panel $p$ shows the synthesized beam ($4\farcs 3\times 3\farcs 4$; $\mathrm{P.A.}=16\degr$).
     The open circles show the field of view at the frequency of $^{13}$CO ($\sim 57\farcs 0$).
 \label{fig:13coch}
 }
\end{figure*}

 Figure~\ref{fig:13coch} shows the channel maps of less optically thick $^{13}$CO ($J=2\text{--}1$) emission. The $^{13}$CO emission was detected in the velocity range from $V_{\rm LSR}=-3.5$ to $3.2\,\mathrm{km\,s^{-1}}$; the blueshifted emission range is smaller than that of the $^{12}$CO emission, which reflects the smaller optical depth of the $^{13}$CO line. At blueshifted channels (panels $j\text{--}l$), the emission components are located south of the protostellar position, which is roughly consistent with the $^{12}$CO emission. The differences in some details may arise from the less optically thick nature of the $^{13}$CO emission. The elongated feature seen in $^{13}$CO emission at the redshifted channels (panels $n$ and $o$) is also seen in the $^{12}$CO emission at the same velocities. On the other hand, at the channels around the systemic velocity (panels $l$ and $m$), where the $^{12}$CO emission may largely be resolved out, we still see the strong $^{13}$CO emission confined to a few features. This is because the optical depth of its $^{13}$CO emission is thinner than that of the $^{12}$CO emission, which allows us to selectively detect compact features. This fact becomes clearer in C$^{18}$O maps below, which is the optically thinnest line of the three.

\begin{figure*}
    \centering
    \includegraphics[bb=0 0 1493 566, width=0.8\hsize]{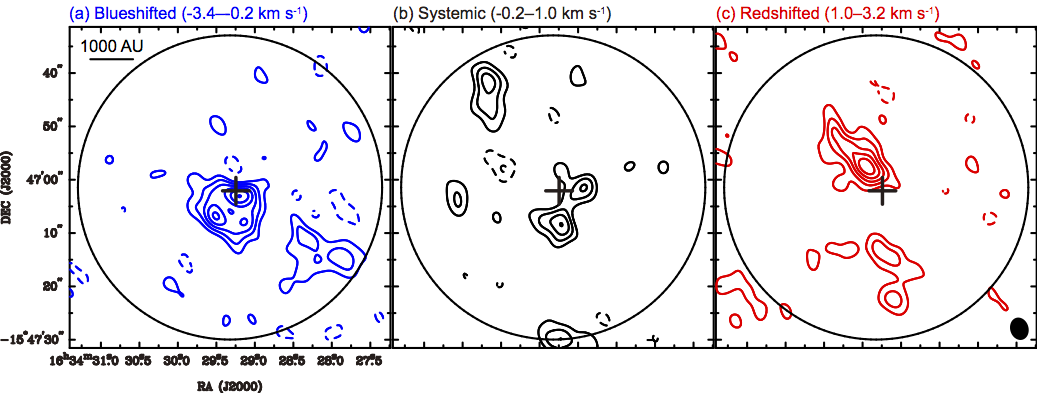}
    \caption{
      Integrated intensity maps of the $^{13}$CO ($J=2\text{--}1$) line emission in L43 observed with the SMA. Each panel shows the integrated intensity map over (a) blueshifted (panels $j\text{--}l$ in Figure~\ref{fig:13coch}), (b) systemic (panel $m$), and (c) redshifted (panels $n$ and $o$) channels. The contour levels are from $3\sigma$ in steps of $2\sigma$, where the $1\sigma$ noise levels are $0.290\,\mathrm{Jy\,beam^{-1}\,km\,s^{-1}}$, $0.167\,\mathrm{Jy\,beam^{-1}\,km\,s^{-1}}$, and $0.237\,\mathrm{Jy\,beam^{-1}\,km\,s^{-1}}$, respectively. The crosses indicate the peak position of the 225~GHz continuum emission in Figure~\ref{fig:cnt}, and the filled ellipse at the bottom right corner in panel (c) shows the synthesized beam ($4\farcs 3\times 3\farcs 4$; $\mathrm{P.A.}=16\degr$). The open circles show the field of view at the frequency of $^{13}$CO ($\sim 57\farcs 0$).
      \label{fig:13comom0}
    }
\end{figure*}
 The integrated intensity maps of the $^{13}$CO ($J=2\text{--}1$) emission are also shown in Figure~\ref{fig:13comom0}, where the integrated velocity ranges are $-3.5\text{--}-0.2\,\mathrm{km\,s^{-1}}$, $-0.2\text{--}1.0\,\mathrm{km\,s^{-1}}$, and $1.0\text{--}3.2\,\mathrm{km\,s^{-1}}$ for blueshifted, systemic, and redshifted velocities, respectively. In the blueshifted velocity, $^{13}$CO emission shows an extended feature with two peaks ($\sim 7\arcsec$ southwest and $\sim 6\arcsec$ southeast). As with the $^{12}$CO emission, the blueshifted components tend to be located south of the protostellar position. In the redshifted velocity, the $^{13}$CO emission also traces the elongated feature identified in the $^{12}$CO map. We interpret that, at these velocity ranges, $^{13}$CO emission traces the blueshifted and redshifted outflows as the $^{12}$CO emission does. Around the systemic velocity, the $^{13}$CO emission has another feature at $\sim 8\arcsec$ south of the protostellar position, which is not visible in the $^{12}$CO emission. The reason why this other feature is visible only in the $^{13}$CO emission is that the $^{13}$CO emission is optically thinner than the $^{12}$CO emission. This feature may be an envelope component, which is resolved out in the $^{12}$CO map.

We thus conclude that the origin of the $^{12}$CO ($J=2\text{--}1$) and $^{13}$CO ($J=2\text{--}1$) emission is mainly the blueshifted and redshifted outflow components. We consider that our data resolve the most central part of the $U$-shaped outflow traced by $^{12}$CO ($J=1\text{--}0$) \citep{lee02}. In the case of the $^{13}$CO emission, the envelope components may be detected around the systemic velocity, and we will confirm this in the following section with the least optically thick C$^{18}$O ($J=2\text{--}1$) emission.

\subsubsection{C$^{18}$O ($J=2\text{--}1$) Emission}\label{sec:c18o}
\begin{figure*}
    \centering
    \includegraphics[bb=0 0 569 549, width=0.8\hsize]{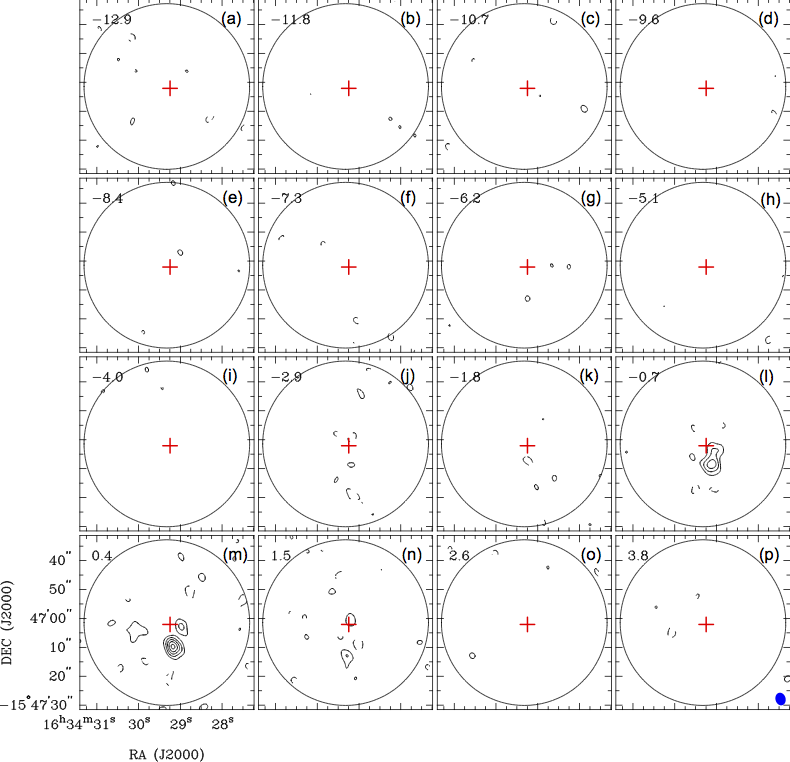}
    \caption{
      Velocity channel maps of the C$^{18}$O ($J=2\text{--}1$) line emission in L43 observed with the SMA.
      The contour levels are from $3\sigma$ in steps of $2\sigma$ ($1\sigma =157\,\mathrm{mJy\,beam^{-1}}$).
      The central velocity of each channel is shown at the top left in units of $\mathrm{km\,s^{-1}}$.
      The crosses indicate the peak position of the 225\,GHz continuum emission in Figure~\ref{fig:cnt}, and the filled ellipse at the bottom right corner in panel $p$ shows the synthesized beam ($4\farcs 3\times 3\farcs 4$; $\text{P.A.}=16\degr$). The open circles show the field of view at the frequency of C$^{18}$O ($\sim 57\farcs 2$).
      \label{fig:C18Ochmap}
    }
\end{figure*}
 Figure~\ref{fig:C18Ochmap} shows the velocity channel maps of the C$^{18}$O ($J=2\text{--}1$) emission, which is detected in three channels around the systemic velocity (panels $l\text{--}n$). It is clear that the line width of the C$^{18}$O emission line is much narrower than those of the C$^{12}$O and $^{13}$CO emission lines. At panel-$m$, which is the closest to the systemic velocity, there are two apparent emission peaks, which we call ``blobs'' here: one at $\sim 7\arcsec$ south of the star (Blob~1) and the other at $\sim 3\arcsec$ west (Blob~2). At blueshifted channel (panel $l$), the emission is elongated from north to south, and seems to consist of two connected features corresponding to blobs. At the redshifted channel (panel $n$), two weak features, probably part of Blob~1 and Blob~2 respectively, are seen. In contrast to the $^{12}$CO and $^{13}$CO emission described in the previous section, the C$^{18}$O emission does not come from either the southern blueshifted component or the northern redshifted one.

\begin{figure*}
    \centering
    \includegraphics[bb=0 0 170 85, width=0.8\hsize]{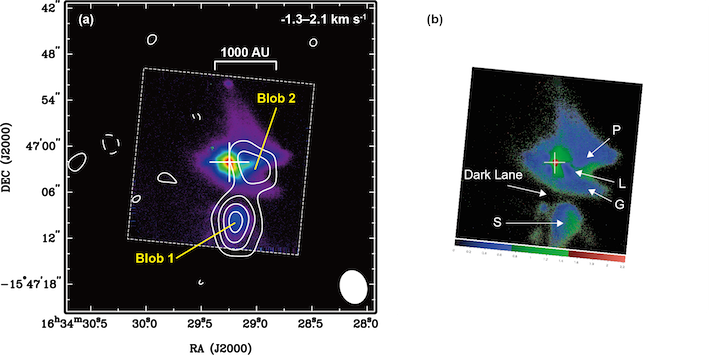}
    \caption{
      (a) Integrated intensity map of the C$^{18}$O ($J=2\text{--}1$) line emission (white contours),
      overlaid on the $K$-band image obtained with the Subaru Telescope \citep{mayama07}.
      The integrated $V_\mathrm{LSR}$ range is indicated at the top right of the panel in units of $\mathrm{km\,s^{-1}}$.
      The C$^{18}$O contour levels are from $3\sigma$ in steps of $2\sigma$ ($1\sigma =0.254\,\mathrm{Jy\,beam^{-1}\,km\,s^{-1}}$). The filled ellipse at the bottom right corner shows the synthesized beam ($4\farcs 4\times 3\farcs 4$; $\mathrm{P.A.}=15\degr$).
      (b) $H-K$ color image obtained with the Subaru Telescope (Fig. 2 in \citealt{mayama07}). P, L, G, and S are the names of components labeled in the reference.
      \label{fig:C18Oblob1}
    }
\end{figure*}
\begin{figure}
    \centering
    \includegraphics[bb=0 0 284 188, width=\hsize]{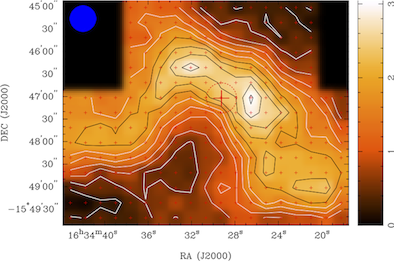}
    \caption{
      Total integrated intensity map of the C$^{18}$O line in L43 obtained with the SMT. The integrated velocity range is from $-1.1\,\mathrm{km\,s^{-1}}$ to $1.9\,\mathrm{km\,s^{-1}}$. The small crosses show the mapped points, and the big cross shows the protostellar position. The filled circle at the top left shows the SMT beam size ($\sim 34.3\arcsec$). The contour levels are from $3\sigma$ in steps of $3\sigma$ ($1\sigma =90\,\mathrm{mK\,km\,s^{-1}}$). The dashed circle indicates the region within $40\arcsec$ from the protostar, where we calculate the LTE mass and virial mass of the protostellar envelope.
      \label{fig:c18osmt}
    }
\end{figure}

 Figure~\ref{fig:C18Oblob1} shows the total integrated intensity map of the C$^{18}$O ($J=2\text{--}1$) emission observed with the SMA and the near-infrared images of L43 taken with the Subaru Telescope \citep{mayama07}. Two C$^{18}$O emission regions, Blobs~1 and 2 mentioned above, are detected above $11\sigma$ and $7\sigma$, respectively. There is no emission component seen toward the protostellar position, where the 225~GHz continuum emission is detected. Previous observations of Class~0 and Class~I protostars such as B335 and L1527 show that there are intense ($\gtrsim 1\,\mathrm{Jy\,beam^{-1}}$) C$^{18}$O emission components associated with the central protostars \citep{jorgensen07,yen10,yen11}. These C$^{18}$O components associated with Class~0 and I protostars have been naturally considered to be molecular envelopes surrounding protostars, indicating that the C$^{18}$O emission is an excellent tracer for protostellar envelopes. Although there is no C$^{18}$O emission component seen toward L43, it does not necessarily mean  there is no C$^{18}$O emission around L43; it is possible there is still C$^{18}$O emission around L43, but most of it was resolved out by SMA. In fact, the C$^{18}$O integrated intensity map obtained with SMT, shown in Figure~\ref{fig:c18osmt}, shows that there is C$^{18}$O emission detectable around L43, although there is no strong peak at the protostellar position of L43. According to the comparison with the spectra toward the center of L43 obtained with SMT and SMA, the missing flux of C$^{18}$O ($J=2\text{--}1$) observed with SMA is estimated to be $92\%$. Importantly, the SMT C$^{18}$O map also shows there is no C$^{18}$O integrated intensity peak at the position of L43 protostar, suggesting that the C$^{18}$O emission is not centrally concentrated at the position of the L43 protostar. This less-concentrated distribution of the C$^{18}$O emission around the L43 protostar is the reason why most of the C$^{18}$O emission was resolved out in the SMA observations.

 The reflection nebulosity associated with L43 is confirmed in near-infrared images \citep[e.g.][]{mayama07}. According to the comparison of the C$^{18}$O emission with the Subaru $K$-band and $H-K$ images, these C$^{18}$O blobs correlate well with the reflection nebulosity. In overlaying the SMA and the Subaru images, we assume that the SMA continuum peak position measured from a 2-dimensional Gaussian fitting coincides with the peak position of the $K$-band image, which represents the protostellar position of L43. In the near-infrared images, Blob~1 corresponds to S, and Blob~2 is located close to L and is sandwiched between P and G. The ``dark lane'' in the near-infrared images corresponds to the valley between blobs in the C$^{18}$O map. Accepting the LTE condition and the optically-thin C$^{18}$O ($J=2\text{--}1$) emission, the integrated intensity in the unit of K km s$^{-1}$ ($\equiv\int T_\mathrm{mb}\ dv$) were converted into the C$^{18}$O column densities as;
\begin{align}
N_\mathrm{mol}&=\frac{8\pi\nu^{3}}{c^{3}}\frac{1}{g_\mathrm{u}A}
\frac{Z({T_\mathrm{ex}})}
{\exp(-E_\mathrm{u}/kT_\mathrm{ex})(\exp(h\nu/kT_\mathrm{ex})-1)} \notag\\
& \qquad\qquad\qquad\qquad\qquad
\times\frac{\int T_\mathrm{mb} dv}{J(T_\mathrm{ex})-J(T_\mathrm{bg})},
\end{align}
where
\begin{align}
J(T)=\frac{h\nu/k}{\exp(h\nu/kT)-1}. 
\end{align}
In the above expressions, $h$ is the Planck constant, $k$ is Boltzmann's constant, $c$ is the speed of light, $\nu$ is the line frequency, $T_\mathrm{ex}$ is the excitation temperature, $T_\mathrm{bg}$ is the background radiation temperature, $A$ is the Einstein $A$-coefficient, $Z({T_\mathrm{ex}})$ is the partition function, $E_\mathrm{u}$ is the rotational energy level of the upper energy state, $g_\mathrm{u}$ is the statistical weight of the upper energy state, and $\int T_\mathrm{mb} dv$ is the integrated line intensity. The C$^{18}$O molecular parameters were taken from \citet{schoier05}. C$^{18}$O column densities are converted into H$_{2}$ column densities ($\equiv N_\mathrm{H_{2}}$) using the C$^{18}$O molecular abundance ($\equiv X_\mathrm{C^{18}O}$) as $N(\mathrm{H_{2}})=N_\mathrm{mol}/X_\mathrm{C^{18}O}$. In accordance with \citet{chen09}, we assumed $X_\mathrm{C^{18}O}=3.1\times 10^{-8}$ taking account of C$^{18}$O depletion factor of $\sim 15$. When $T_\mathrm{ex}$ is 10~K, $N(\mathrm{H_{2}})$ at the positions of Blobs 1 and 2 is estimated to be $1.0\times 10^{23}\,\mathrm{cm^{-2}}$ and $7.2\times 10^{22}\,\mathrm{cm^{-2}}$, respectively. On the other hand, $N_\mathrm{H_{2}}$ is expressed using color excess by dust particles as
\begin{align}
 N_\mathrm{H} = N(\mathrm{H}) + 2 N(\mathrm{H_{2}}) = 4.77\times 10^{21} E(B-V)~\mathrm{cm^{-2}}.
\end{align}
Assuming $A_\mathrm{V}/E(\mathrm{B - V})\sim 3$ and $A_\mathrm{K}/A_\mathrm{V}=0.15$ \citep{scheffler87}, derived column densities of molecular hydrogen translate into dust extinction at $K$-band as follows:
\begin{align}
 A_\mathrm{K} = 0.189\left(\frac{N(\mathrm{H_{2}})}{1.0\times 10^{21}~\mathrm{cm^{-2}}}\right).
\end{align}
Derived physical parameters of Blobs 1 and 2 are summarized in Table~\ref{tab:blobs}.
\begin{deluxetable}{lccccccc}
 \tabletypesize{\scriptsize}
 \tablecaption{
 Physical properties of the C$^{18}$O blobs
 \label{tab:blobs}
 }
 \tablewidth{0pt}
 \tablehead{
 &
 \colhead{($\Delta\alpha$,$\Delta\delta$)\tablenotemark{a}} &
 \colhead{$\theta_\mathrm{FWHM}$\tablenotemark{b}} &
 \colhead{$N_\mathrm{H_{2}}$\tablenotemark{c}} &
 \colhead{$A_\mathrm{K}$\tablenotemark{d}} \\
 \colhead{Component} &
 \colhead{(arcsec)} &
 \colhead{(arcsec)} &
 \colhead{($\mathrm{cm^{-2}}$)} &
 \colhead{(mag)}
 }
 \startdata
 Blob~1 &
 ($-0\farcs 9$, $-8\farcs 0$) &
 $6\farcs 5\times 4\farcs 8$ &
 $1.0\times 10^{23}$ &
 $19.6$
 \\
 Blob~2 &
 ($-3\farcs 2$, $-0\farcs 4$) &
 $6\farcs 3\times 4\farcs 8$ &
 $7.2\times 10^{22}$ &
 $13.6$
 \enddata
 \tablenotetext{a}{Offset position from the central protostar.}
 \tablenotetext{b}{FWHM size of the 2-dimensional Gaussian fitting (major and minor axes).}
 \tablenotetext{c}{Peak column density of molecular hydrogen toward each component.}
 \tablenotetext{d}{Extinction at $K$-band.}
\end{deluxetable}

The dust extinction of $> 10\,\mathrm{mag}$ at $K$-band allows us to observe scattered light at the positions of the blobs. In the case that the blobs are located on the near side with respect to the position of the central star from the observer's point of view, however, the scattered light would be totally obscured by the blobs themselves. Therefore, the blobs would be located on the far side, and we would observe the backward scattering at the blobs. The $H-K$ color image can also be interpreted in the same manner. The west sides of the blobs, which correspond to the far sides with respect to the central protostar, spatially correspond to the green regions in the $H-K$ image. This is because shorter waves are scattered at the near sides, which correspond to the east sides. Therefore, only longer waves can reach and are scattered at the far sides. Our comparison of these images suggests that the envelope material is so transparent that we can observe the near-infrared emission from the star. It also indicates that the dark lane seen in near-infrared images is not a shadow of material such as a circumstellar disk, but a gap in the material. More details of the nature of the C$^{18}$O envelope in L43 will be discussed later.

 \section{Discussion}\label{sec:discuss}
  \subsection{Dust Opacity Index $\beta$}\label{sec:beta}
  The slope of the dust thermal emission provide us the information about the nature of dust particles through dust opacity indices. If we assume a single-temperature optically thin dust emission, then the flux density is expressed as
  \begin{align}
      F_{\nu} \propto \kappa_{\nu}B_{\nu}(T_\mathrm{d}) \propto \nu^{\alpha},
  \end{align}
  where $B_{\nu}(T_\mathrm{d})$ is the Planck function, $T_\mathrm{d}$ is the dust temperature, and $\kappa_{\nu}$ is the dust opacity at a frequency $\nu$. The dust opacity $\kappa_{\nu}$ is usually expressed as $\kappa_{\nu} \propto \nu^{\beta}$ at radio wavelengths. Therefore, the dust opacity index $\beta$ can be estimated from a power-law fitting to $\kappa_{\nu}\propto F_{\nu}/B_{\nu}(T_\mathrm{d})$. We performed a $\chi^{2}$ fitting to the single-dish and interferometric measurements separately because of the gap between millimeter measurements using interferometer and far-infrared ones using the single-dish measurement mentioned above. As far-infrared measurements, two flux densities at $450\,\mathrm{\mu m}$ and $850\,\mathrm{\mu m}$ marked in blue in Figure~\ref{fig:sed} are used, since these measurements were obtained simultaneously with SCUBA, and their relative intensities are reliable. The power-law indices $\beta$ are derived to be $2.8$ for the far-infrared measurements and $0.46\pm 0.47$ for millimeter measurements with $T_\mathrm{d}=10\,\mathrm{K}$. Since there are only two measurements in far-infrared wavelengths, no error in $\beta$ arises in fitting. The assumption of higher temperature decreases the value of $\beta$. If we assume $T_\mathrm{d}=30\,\mathrm{K}$, $\beta$ would be $1.6$ and $0.17\pm 0.49$ for far-infrared and millimeter measurements, respectively. The difference in the beam sizes may have an influence on the flux measurements at the millimeter wavelengths because of the missing flux. We roughly estimated this effect based on the minimum baseline length of each observation and our envelope models (see \S~\ref{sec:dust} for the details of our models). The measured flux density with the SMA is $\sim 50\%$ lower than those measured with the larger beams of the OVRO observations in the worst case. If we take this into consideration, $\beta$ at the millimeter wavelengths is estimated to be $1.0\pm 0.3$ at most.

  The value of $\beta$ derived from millimeter measurements is similar to those of protoplanetary disks around T Tauri stars \citep{beckwith91,mannings94,andrews07} rather than that of the interstellar value of $\sim 1.7$ \citep{hildebrand83,weingartner01} and those adopted for low-mass protostellar sources (1--1.7) \citep{ohashi96,ossenkopf94}. These $\beta$ values smaller than 1 can be interpreted in the context of dust growth \citep{dalessio01,draine06}, which is possible in high-density environments such as protoplanetary disks around T~Tauri stars. The millimeter measurements using interferometry toward L43, which show the compact feature, suggest dust growth in the vicinity of the central star ($< 500\,\mathrm{AU}$). Recent observations have also revealed that SEDs of Class 0 and Class I sources show such small $\beta$ values \citep{kwon09,scaife12}. Our result supports dust growth at an early stage of star formation.
  
  The total (gas + dust) mass of the compact circumstellar component ($\equiv M_\mathrm{d}$) can also be estimated from the continuum flux density as:
  \begin{align}
   M_\mathrm{d}=\frac{F_{\nu}d^{2}}{\kappa_{\nu} B_{\nu}(T_\mathrm{d})},
    \label{eq:mdust}
  \end{align}
  where $d$ is the distance to the target. On the assumption of $\kappa_{0}=0.1\,\mathrm{cm^{2}\,g^{-1}}$ at $1000\,\mathrm{GHz}$ \citep{beckwith90}, $\beta=0.46$ derived above for millimeter measurements with $T_\mathrm{d}=10\,\mathrm{K}$ yields $\kappa_{\nu}=5.1\times 10^{-2}\,\mathrm{cm^{2}\,g^{-1}}$ at $225\,\mathrm{GHz}$. With this dust opacity, $M_\mathrm{d}$ was estimated to be $4.0\times 10^{-3}\,\mathrm{M_{\odot}}$. If a higher temperature of 30~K is adopted, $M_\mathrm{d}$ increases to $5.9\times 10^{-4}\,\mathrm{M_{\odot}}$ with $\kappa_{\nu}=7.7\times 10^{-2}\,\mathrm{cm^{2}\,g^{-1}}$ at $225\,\mathrm{GHz}$. These results show that $M_\mathrm{d}$ is probably around $10^{-3}\,\mathrm{M_{\odot}}$, although the absolute value largely depends on the assumption of the dust opacity and temperature. We discuss the origin of this dust thermal emission in the following subsection.

  \subsection{Origin of Dust Continuum Emission}\label{sec:dust}
  As was shown earlier, weak continuum emission was detected at the stellar position. The appearance of the emission is compact with some elongation, which could suggest that the emission arises from a compact circumstellar disk associated with the central star. The compactness of the emission, however, might be artificially created by our SMA observations resolving out extended structures. It is also noteworthy that typical Class~0 and Class~I protostars are often associated with dust continuum emission arising from an envelope surrounding the protostar \citep[e.g.][]{arce06,jorgensen07}.

  In order to investigate whether the continuum emission arises from the innermost part of the extended envelope, we used a simple model of a spherical core to see whether we could reproduce the results obtained with SMA at 225~GHz. In the model, the intensity distribution of continuum emission at $225\,\mathrm{GHz}$ was assumed to be a power-law, $I(r)\propto r^{-q}$.
The power-law index, $q$, is estimated to be $1.1\pm 0.39$ from the $450\,\mathrm{\mu m}$ data \citep{shirley00}. The total flux density of the continuum emission at $225\,\mathrm{GHz}$ within the core is estimated to be $274\,\mathrm{mJy}$ by extrapolation from the far-infrared measurements at $450\,\mathrm{\mu m}$ and $850\,\mathrm{\mu m}$ (see the blue dashed line in Figure~\ref{fig:sed}). To avoid divergence at $r=0$, we assumed that the innermost part of the distribution has a hole where the intensity falls to zero. We tried various hole sizes from $50\,\mathrm{AU}$ to $800\,\mathrm{AU}$ in radius. With these models, observations using the actual SMA antenna configuration were simulated with the CASA Simulator.

\begin{figure}
    \centering
    \includegraphics[bb=0 0 428 626, width=\hsize]{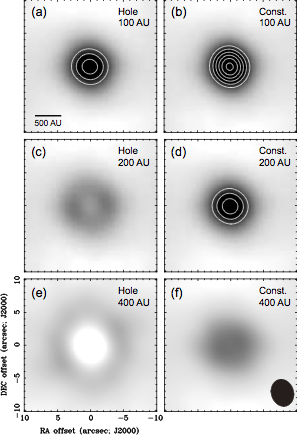}
    \caption{Examples of $225\,\mathrm{GHz}$ continuum maps obtained with our simulated observations.
      The radius of the hole or constant region is shown at the top right in each panel.
      The filled ellipse at the bottom right in panel $f$ is the synthesized beam.
      The contour levels and the size of maps are the same as Figure~\ref{fig:cnt}.
      \label{fig:qobs}}
\end{figure}

  The results of our simulated observations are summarized in Table~\ref{tab:qobs} and Fugure~\ref{fig:qobs}. For the model with a hole of $50\,\mathrm{AU}$ in radius, the expected flux density of $23.6\,\mathrm{mJy}$ is nearly twice as large as that of the actual observations. The expected flux density decreases as the radius of the hole becomes larger, and agrees with the observations when the radius is between $100\,\mathrm{AU}$ and $200\,\mathrm{AU}$. Note that the central hole is marginally resolved in the resultant image with a hole of $200\,\mathrm{AU}$ in radius. Models with larger holes have smaller flux densities than the observations, and also show images with holes, which is inconsistent with the observed image.
\begin{deluxetable}{ccc}
    \tabletypesize{\scriptsize}
    \tablecaption{
      Results of our simulated observations
      \label{tab:qobs}
    }
    \tablewidth{0pt}
    \tablehead{
      \colhead{Radius\tablenotemark{a}} & \multicolumn{2}{c}{Expected Flux ($\mathrm{mJy}$)} \\
      \colhead{($\mathrm{AU}$)} & \multicolumn{1}{c}{Hole} & \multicolumn{1}{c}{Constant}
    }
    \startdata
    50 & 37.9 & 41.4 \\
    100 & 30.3 & 38.1 \\
    200 & 19.1 & 30.0 \\
    400 & 8.6 & 19.4 \\
    800 & ...\tablenotemark{b} & ...\tablenotemark{b}
    \enddata
    \tablenotetext{a}{Radius for an inner hole or a constant intensity region.}
    \tablenotetext{b}{Undetected.}
\end{deluxetable}
  
  We also tried models with constant intensities instead of null intensities within certain radii. As shown in the Table 6, a model with a constant intensity within a given radius produces a larger flux density than the model with a hole having the same radius. In order for a model with a constant intensity to produce a flux density consistent with the observations, the radius for a constant intensity should be between $200\,\mathrm{AU}$ and $400\,\mathrm{AU}$. These simulations suggest that the observations can be explained by assuming an envelope with a hole or a constant intensity region within a few hundred AU in radius.

  Note that when the radius of a hole or a constant intensity region becomes more than $800\,\mathrm{AU}$, emission from the envelope is undetectable with the present SMA observations. In such cases, the detected continuum emission can be still explained as that arising from a compact circumstellar disk associated with the central star.
  
  In the cases we discussed above, it is necessary for the innermost part of the envelope to have a hole or a constant intensity region. This suggests that the innermost envelope has less material than an ordinary envelope. The smaller amount of material in the innermost envelope around the L43 protostar is consistent with the results obtained from the comparison between C$^{18}$O map and the NIR scattering image in \S~\ref{sec:c18o}.

\subsection{Disappearance of Protostellar Envelope}\label{sec:env}
In the previous section, it was suggested that the innermost envelope around L43 protostar seem to have a less amount of material. This may imply that the innermost envelope disappears for some reason. We now discuss this possibility in detail.

In the previous section, we used our results obtained with the SMA, which is sensitive to the innermost envelope, but is missing flux for larger-scale structures. In this section, we use single-dish data to tell us the overall distributions of the envelope. In order to investigate the envelope distribution, the intensity distribution of continuum emission is examined here.

\begin{figure}
    \centering
    \includegraphics[bb=0 0 576 432, width=\hsize]{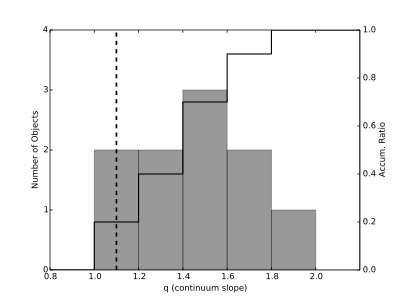}
    \caption{Histogram of the power-law index of the intensity distribution at $450\,\mathrm{\mu m}$, $q$, in protostars (left axis) and its cumulative distribution (right axis). The measurements are taken from \citet{shirley00}. The vertical dashed line indicates the value for L43 ($q=1.1$).
      \label{fig:slope}}
\end{figure}
\citet{shirley00} observed $450\,\mathrm{\mu m}$ continuum emission around various Class 0/I protostars, including the L43 protostar, and fit the intensity distributions as a function of the distance from the protostars with power-laws ($I_{r}\propto r^{-q}$). The power-law index for the envelope around the L43 protostar was estimated to be $1.1\pm 0.39$, as was explained in the previous section. As demonstrated in Figure~\ref{fig:slope}, where the histogram of the power-law index of the intensity distributions for all the Class 0/I protostars observed by \citeauthor{shirley00} is shown, the power-law index for the envelope around the L43 protostar is smaller than those for other protostars. This suggests that the envelope around the L43 protostar is less concentrated compared with those around other protostars. This lower concentration may be a result of the envelope's disappearance.

Since the envelope seems to be disappearing, one may consider that it is not gravitationally bound any more.
In order to investigate whether the envelope is gravitationally bound, we compare the H$_{2}$ gas mass within $40\arcsec$ from the position of the central protostar, where dust continuum at $850\,\mathrm{\mu m}$ shows a clear condensation, with the virial mass.
The H$_{2}$ gas mass of the envelope is estimated to be $1.5\,\mathrm{M_{\odot}}$ for $T=10\,\mathrm{K}$ from the flux density at $850\,\mathrm{\mu m}$ using equation~\eqref{eq:mdust}, where $\kappa_{\nu}=5.4\times 10^{-3}\,\mathrm{cm^{2}\,g^{-1}}$ at $850\,\mathrm{\mu m}$ with $\beta=2.8$.
On the other hand, the virial mass ($\equiv M_\mathrm{vir}$) is estimated from the line profile of C$^{18}$O ($J=2\text{--}1$) emission obtained with SMT as follows;
\begin{align}
 M_\mathrm{vir}=\frac{5DC_\mathrm{eff}^{2}}{2G},
\end{align}
where $G$ is the gravitational constant, $D$ is the size of the object.
$C_\mathrm{eff}$ is the effective sound speed of gas as
\begin{align}
C^{2}_\mathrm{eff}=\frac{\Delta V^{2}}{8\ln2},
\end{align}
where $\Delta V$ denotes the FWHM line width of the C$^{18}$O ($J=2\text{--}1$) emission.
\begin{figure}
    \centering
    \includegraphics[bb=0 0 576 432, width=\hsize]{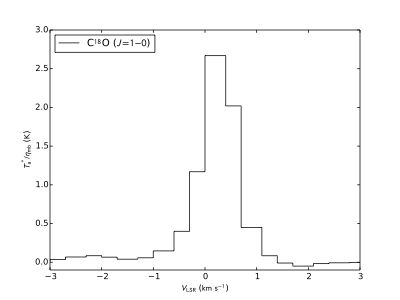}
    \caption{
      Line profile of the C$^{18}$O ($J=2\text{--}1$) emission observed with the SMT within $40\arcsec$ from the central protostar, where we estimated the virial mass of the protostellar envelope. The velocity resolution is $\sim 0.34\,\mathrm{km\,s^{-1}}$.
      \label{fig:c18osp}
    }
\end{figure}
We measure the FWHM line width of the C$^{18}$O profile in Figure~\ref{fig:c18osp} to be $\sim 0.59\,\mathrm{km\,s^{-1}}$ with a Gaussian fitting.
The derived virial mass is $\sim 1.0\,\mathrm{M_{\odot}}$. Hence, if we take into account the central protostellar mass of $\sim 0.5\,\mathrm{M_{\odot}}$ \citep{levreault88}, the envelope mass within $40\arcsec$ is comparable with the virial mass, suggesting that the protostellar envelope around L43 is likely gravitationally bound.

At first glance, this result seems to be inconsistent with our idea that the envelope of L43 has been disappearing. However, we should note that the envelope can disappear when its material accretes onto the central star and the disk system. Since the protostar in L43 is still associated with a strong molecular outflow, it is naturally considered that the material in the envelope is still collecting onto the central star and disk system, and this process consumes the material in the envelope. We suggest that the envelope around L43 is disappearing through a process of accretion, at least in the close vicinity of the protostar.

\section{Summary}
We present SMA data of L43, a transitional protostellar object from the Class I to T-Tauri stage, in the $^{12}$CO ($J=2\text{--}1$), $^{13}$CO ($J=2\text{--}1$), and C$^{18}$O ($J=2\text{--}1$) emission lines, and the 225~GHz continuum emission, together with the new single-dish observations with the SMT in the C$^{18}$O ($J=2\text{--}1$) line. The main results are summarized as follows:
\begin{enumerate}
      \item The $225\,\mathrm{GHz}$ continuum emission shows a weak ($\sim 23.6\pm 2.4\,\mathrm{mJy}$), compact ($< 1000\,\mathrm{AU}$) component associated with the central star. Our simulated observations show the observed continuum emission can be explained by the dust thermal emission from the envelope, which has a hole or a constant intensity region within a few hundred AU of the central protostar. This suggests two possibilities: the disappearance of the envelope or a lower distribution of envelope material over a small-scale region. The dust opacity index $\beta$ ($\sim 0.46$) estimated from the millimeter interferometric observations is much smaller than the typical value of the interstellar medium ($\sim 1.7$) and low-mass protostellar sources ($\sim 1\text{--}1.7$), suggesting possible growth of dust grains.
      \item The $^{12}$CO ($J=2\text{--}1$) and $^{13}$CO ($J=2\text{--}1$) emission exhibit blueshifted outflows toward the south of the central star, and redshifted outflows toward the north. The SMA observations resolve the innermost component of the blueshifted $U$-shaped outflow obtained with previous interferometric mosaicking observations. The direction of these small-scale outflow features, i.e. blueshifted toward the south and redshifted toward the north, is roughly consistent with the direction of the large-scale ($> 50,000\,\mathrm{AU}$) outflow.
      \item The SMA C$^{18}$O map shows two molecular blobs, which correspond to the reflection nebulosity seen in the near-infrared images. The dust extinctions at the positions of blobs ($> 10\,\mathrm{mag}$) suggest that the origin of the $K$-band reflection is backward scattering at the blobs.
      \item The envelope around the L43 protostar shows an intensity distribution with a power-law index of $1.1\pm 0.39$, which is shallower than other protostars, suggesting a lower concentration of material in the envelope around the L43 protostar.
\end{enumerate}
The visible scattering features should result from the optical thinness of the envelope material in L43, which is consistent with the less-concentrated distribution in the continuum emission compared to other protostars. Such a low-concentration distribution of the envelope could be due to its disappearance. Because the mass of the envelope ($\sim 1.5~\text{M$_{\odot}$}$) $+$ protostar ($\sim 0.5\,\mathrm{M_{\odot}}$) is comparable with the virial mass of $M_\mathrm{vir}=1.0\,\mathrm{M_{\odot}}$ within $40\arcsec$ from the protostellar position, the protostellar envelope around L43 is likely gravitationally bound. We conclude that the protostellar envelope (at least in close proximity to the central protostar) is disappearing due to the consumption of envelope material through mass accretion onto the central star and disk system. 


 
 
 \bibliographystyle{yahapj}
 \bibliography{koyamatsu}

\begin{thebibliography}{44}
\providecommand\natexlab[1]{#1}
\providecommand\JournalTitle[1]{#1}

\bibitem[{{Andre} \& {Montmerle}(1994)}]{andre94}
{Andre}, P., \& {Montmerle}, T. 1994,
  \href{http://dx.doi.org/10.1086/173608}{\JournalTitle{\apj}, 420, 837}

\bibitem[{{Andre} {et~al.}(2000){Andre}, {Ward-Thompson}, \&
  {Barsony}}]{andre00}
{Andre}, P., {Ward-Thompson}, D., \& {Barsony}, M. 2000,
  \JournalTitle{Protostars and Planets IV}, 59

\bibitem[{{Andrews} \& {Williams}(2005)}]{andrews05}
{Andrews}, S.~M., \& {Williams}, J.~P. 2005,
  \href{http://dx.doi.org/10.1086/432712}{\JournalTitle{\apj}, 631, 1134}

\bibitem[{{Andrews} \& {Williams}(2007)}]{andrews07}
---. 2007, \href{http://dx.doi.org/10.1086/522885}{\JournalTitle{\apj}, 671,
  1800}

\bibitem[{{Anglada} \& {Rodr{\'{\i}}guez}(2002)}]{anglada02}
{Anglada}, G., \& {Rodr{\'{\i}}guez}, L.~F. 2002, \JournalTitle{RevMexAA}, 38,
  13

\bibitem[{{Arce} \& {Sargent}(2006)}]{arce06}
{Arce}, H.~G., \& {Sargent}, A.~I. 2006,
  \href{http://dx.doi.org/10.1086/505104}{\JournalTitle{\apj}, 646, 1070}

\bibitem[{{Beckwith} \& {Sargent}(1991)}]{beckwith91}
{Beckwith}, S.~V.~W., \& {Sargent}, A.~I. 1991,
  \href{http://dx.doi.org/10.1086/170646}{\JournalTitle{\apj}, 381, 250}

\bibitem[{{Beckwith} {et~al.}(1990){Beckwith}, {Sargent}, {Chini}, \&
  {Guesten}}]{beckwith90}
{Beckwith}, S.~V.~W., {Sargent}, A.~I., {Chini}, R.~S., \& {Guesten}, R. 1990,
  \href{http://dx.doi.org/10.1086/115385}{\JournalTitle{\aj}, 99, 924}

\bibitem[{{Chen} {et~al.}(1995){Chen}, {Myers}, {Ladd}, \& {Wood}}]{chen95}
{Chen}, H., {Myers}, P.~C., {Ladd}, E.~F., \& {Wood}, D.~O.~S. 1995,
  \href{http://dx.doi.org/10.1086/175703}{\JournalTitle{\apj}, 445, 377}

\bibitem[{{Chen} {et~al.}(2009){Chen}, {Evans}, {Lee}, \& {Bourke}}]{chen09}
{Chen}, J.-H., {Evans}, II, N.~J., {Lee}, J.-E., \& {Bourke}, T.~L. 2009,
  \href{http://dx.doi.org/10.1088/0004-637X/705/2/1160}{\JournalTitle{\apj},
  705, 1160}

\bibitem[{{D'Alessio} {et~al.}(2001){D'Alessio}, {Calvet}, \&
  {Hartmann}}]{dalessio01}
{D'Alessio}, P., {Calvet}, N., \& {Hartmann}, L. 2001,
  \href{http://dx.doi.org/10.1086/320655}{\JournalTitle{\apj}, 553, 321}

\bibitem[{{de Geus} {et~al.}(1990){de Geus}, {Bronfman}, \&
  {Thaddeus}}]{de-geus90}
{de Geus}, E.~J., {Bronfman}, L., \& {Thaddeus}, P. 1990, \JournalTitle{\aap},
  231, 137

\bibitem[{{Draine}(2006)}]{draine06}
{Draine}, B.~T. 2006,
  \href{http://dx.doi.org/10.1086/498130}{\JournalTitle{\apj}, 636, 1114}

\bibitem[{{Greene} {et~al.}(1994){Greene}, {Wilking}, {Andre}, {Young}, \&
  {Lada}}]{greene94}
{Greene}, T.~P., {Wilking}, B.~A., {Andre}, P., {Young}, E.~T., \& {Lada},
  C.~J. 1994, \href{http://dx.doi.org/10.1086/174763}{\JournalTitle{\apj}, 434,
  614}

\bibitem[{{Hayashi} {et~al.}(1993){Hayashi}, {Ohashi}, \& {Miyama}}]{hayashi93}
{Hayashi}, M., {Ohashi}, N., \& {Miyama}, S.~M. 1993,
  \href{http://dx.doi.org/10.1086/187119}{\JournalTitle{\apjl}, 418, L71}

\bibitem[{{Heyer} {et~al.}(1990){Heyer}, {Ladd}, {Myers}, \&
  {Campbell}}]{heyer90}
{Heyer}, M.~H., {Ladd}, E.~F., {Myers}, P.~C., \& {Campbell}, B. 1990,
  \href{http://dx.doi.org/10.1086/115441}{\JournalTitle{\aj}, 99, 1585}

\bibitem[{{Hildebrand}(1983)}]{hildebrand83}
{Hildebrand}, R.~H. 1983, \JournalTitle{\qjras}, 24, 267

\bibitem[{{Ho} {et~al.}(2004){Ho}, {Moran}, \& {Lo}}]{ho04}
{Ho}, P.~T.~P., {Moran}, J.~M., \& {Lo}, K.~Y. 2004,
  \href{http://dx.doi.org/10.1086/423245}{\JournalTitle{\apjl}, 616, L1}

\bibitem[{{J{\o}rgensen} {et~al.}(2007){J{\o}rgensen}, {Bourke}, {Myers}, {Di
  Francesco}, {van Dishoeck}, {Lee}, {Ohashi}, {Sch{\"o}ier}, {Takakuwa},
  {Wilner}, \& {Zhang}}]{jorgensen07}
{J{\o}rgensen}, J.~K., {Bourke}, T.~L., {Myers}, P.~C., {et~al.} 2007,
  \href{http://dx.doi.org/10.1086/512230}{\JournalTitle{\apj}, 659, 479}

\bibitem[{{Kitamura} {et~al.}(1996){Kitamura}, {Kawabe}, \&
  {Saito}}]{kitamura96}
{Kitamura}, Y., {Kawabe}, R., \& {Saito}, M. 1996,
  \href{http://dx.doi.org/10.1086/176728}{\JournalTitle{\apj}, 457, 277}

\bibitem[{{Kwon} {et~al.}(2009){Kwon}, {Looney}, {Mundy}, {Chiang}, \&
  {Kemball}}]{kwon09}
{Kwon}, W., {Looney}, L.~W., {Mundy}, L.~G., {Chiang}, H.-F., \& {Kemball},
  A.~J. 2009,
  \href{http://dx.doi.org/10.1088/0004-637X/696/1/841}{\JournalTitle{\apj},
  696, 841}

\bibitem[{{Lee} \& {Ho}(2005)}]{lee05}
{Lee}, C.-F., \& {Ho}, P.~T.~P. 2005,
  \href{http://dx.doi.org/10.1086/429535}{\JournalTitle{\apj}, 624, 841}

\bibitem[{{Lee} {et~al.}(2002){Lee}, {Mundy}, {Stone}, \& {Ostriker}}]{lee02}
{Lee}, C.-F., {Mundy}, L.~G., {Stone}, J.~M., \& {Ostriker}, E.~C. 2002,
  \href{http://dx.doi.org/10.1086/341540}{\JournalTitle{\apj}, 576, 294}

\bibitem[{{Levreault}(1988)}]{levreault88}
{Levreault}, R.~M. 1988,
  \href{http://dx.doi.org/10.1086/166520}{\JournalTitle{\apj}, 330, 897}

\bibitem[{{Mannings} \& {Emerson}(1994)}]{mannings94}
{Mannings}, V., \& {Emerson}, J.~P. 1994, \JournalTitle{\mnras}, 267, 361

\bibitem[{{Mayama} {et~al.}(2007){Mayama}, {Tamura}, {Hayashi}, {Itoh},
  {Ishii}, {Fukagawa}, {Hayashi}, {Oasa}, \& {Kudo}}]{mayama07}
{Mayama}, S., {Tamura}, M., {Hayashi}, M., {et~al.} 2007, \JournalTitle{\pasj},
  59, 1153

\bibitem[{{Momose} {et~al.}(1996){Momose}, {Ohashi}, {Kawabe}, {Hayashi}, \&
  {Nakano}}]{momose96}
{Momose}, M., {Ohashi}, N., {Kawabe}, R., {Hayashi}, M., \& {Nakano}, T. 1996,
  \href{http://dx.doi.org/10.1086/177925}{\JournalTitle{\apj}, 470, 1001}

\bibitem[{{Myers} {et~al.}(2000){Myers}, {Evans}, \& {Ohashi}}]{myers00}
{Myers}, P.~C., {Evans}, II, N.~J., \& {Ohashi}, N. 2000,
  \JournalTitle{Protostars and Planets IV}, 217

\bibitem[{{Myers} {et~al.}(1987){Myers}, {Fuller}, {Mathieu}, {Beichman},
  {Benson}, {Schild}, \& {Emerson}}]{myers87}
{Myers}, P.~C., {Fuller}, G.~A., {Mathieu}, R.~D., {et~al.} 1987,
  \href{http://dx.doi.org/10.1086/165458}{\JournalTitle{\apj}, 319, 340}

\bibitem[{{Nakano} {et~al.}(1995){Nakano}, {Hasegawa}, \& {Norman}}]{nakano95}
{Nakano}, T., {Hasegawa}, T., \& {Norman}, C. 1995,
  \href{http://dx.doi.org/10.1086/176130}{\JournalTitle{\apj}, 450, 183}

\bibitem[{{Ohashi} {et~al.}(1996){Ohashi}, {Hayashi}, {Kawabe}, \&
  {Ishiguro}}]{ohashi96}
{Ohashi}, N., {Hayashi}, M., {Kawabe}, R., \& {Ishiguro}, M. 1996,
  \href{http://dx.doi.org/10.1086/177512}{\JournalTitle{\apj}, 466, 317}

\bibitem[{{Ossenkopf} \& {Henning}(1994)}]{ossenkopf94}
{Ossenkopf}, V., \& {Henning}, T. 1994, \JournalTitle{\aap}, 291, 943

\bibitem[{{Sault} {et~al.}(1995){Sault}, {Teuben}, \& {Wright}}]{sault95}
{Sault}, R.~J., {Teuben}, P.~J., \& {Wright}, M.~C.~H. 1995, in ASP Conf. Ser.
  77, Astronomical Data Analysis Software and Systems IV, ed. R.~A. {Shaw},
  H.~E. {Payne}, \& J.~J.~E. {Hayes (San Francisco, CA: ASP)}, 433

\bibitem[{{Scaife} {et~al.}(2012){Scaife}, {Buckle}, {Ainsworth}, {Davies},
  {Franzen}, {Grainge}, {Hobson}, {Hurley-Walker}, {Lasenby}, {Olamaie},
  {Perrott}, {Pooley}, {Ray}, {Richer}, {Rodr{\'{\i}}guez-Gonz{\'a}lvez},
  {Saunders}, {Schammel}, {Scott}, {Shimwell}, {Titterington}, \&
  {Waldram}}]{scaife12}
{Scaife}, A.~M.~M., {Buckle}, J.~V., {Ainsworth}, R.~E., {et~al.} 2012,
  \href{http://dx.doi.org/10.1111/j.1365-2966.2011.20254.x}{\JournalTitle{\mnras},
  420, 3334}

\bibitem[{{Scarrott} {et~al.}(1993){Scarrott}, {Draper}, \&
  {Tadhunter}}]{scarrott93}
{Scarrott}, S.~M., {Draper}, P.~W., \& {Tadhunter}, C.~N. 1993,
  \JournalTitle{\mnras}, 262, 306

\bibitem[{{Scheffler} \& {Elsaesser}(1987)}]{scheffler87}
{Scheffler}, H., \& {Elsaesser}, H. 1987, {Physics of the galaxy and
  interstellar matter}

\bibitem[{{Sch{\"o}ier} {et~al.}(2005){Sch{\"o}ier}, {van der Tak}, {van
  Dishoeck}, \& {Black}}]{schoier05}
{Sch{\"o}ier}, F.~L., {van der Tak}, F.~F.~S., {van Dishoeck}, E.~F., \&
  {Black}, J.~H. 2005,
  \href{http://dx.doi.org/10.1051/0004-6361:20041729}{\JournalTitle{\aap}, 432,
  369}

\bibitem[{{Scoville} {et~al.}(1993){Scoville}, {Carlstrom}, {Chandler},
  {Phillips}, {Scott}, {Tilanus}, \& {Wang}}]{scoville93}
{Scoville}, N.~Z., {Carlstrom}, J.~E., {Chandler}, C.~J., {et~al.} 1993,
  \href{http://dx.doi.org/10.1086/133332}{\JournalTitle{\pasp}, 105, 1482}

\bibitem[{{Shirley} {et~al.}(2000){Shirley}, {Evans}, {Rawlings}, \&
  {Gregersen}}]{shirley00}
{Shirley}, Y.~L., {Evans}, II, N.~J., {Rawlings}, J.~M.~C., \& {Gregersen},
  E.~M. 2000, \href{http://dx.doi.org/10.1086/317358}{\JournalTitle{\apjs},
  131, 249}

\bibitem[{{Terebey} {et~al.}(1993){Terebey}, {Chandler}, \&
  {Andre}}]{terebey93}
{Terebey}, S., {Chandler}, C.~J., \& {Andre}, P. 1993,
  \href{http://dx.doi.org/10.1086/173121}{\JournalTitle{\apj}, 414, 759}

\bibitem[{{Weingartner} \& {Draine}(2001)}]{weingartner01}
{Weingartner}, J.~C., \& {Draine}, B.~T. 2001,
  \href{http://dx.doi.org/10.1086/318651}{\JournalTitle{\apj}, 548, 296}

\bibitem[{{Wilner} \& {Welch}(1994)}]{wilner94}
{Wilner}, D.~J., \& {Welch}, W.~J. 1994,
  \href{http://dx.doi.org/10.1086/174195}{\JournalTitle{\apj}, 427, 898}

\bibitem[{{Yen} {et~al.}(2010){Yen}, {Takakuwa}, \& {Ohashi}}]{yen10}
{Yen}, H.-W., {Takakuwa}, S., \& {Ohashi}, N. 2010,
  \href{http://dx.doi.org/10.1088/0004-637X/710/2/1786}{\JournalTitle{\apj},
  710, 1786}

\bibitem[{{Yen} {et~al.}(2011){Yen}, {Takakuwa}, \& {Ohashi}}]{yen11}
---. 2011,
  \href{http://dx.doi.org/10.1088/0004-637X/742/1/57}{\JournalTitle{\apj}, 742,
  57}

\end{thebibliography}
 
\end{document}